\documentclass[conference]{IEEEtran}
\usepackage{graphicx, subfigure}
\graphicspath{{figures/}}

\usepackage{url}
\usepackage{xspace}
\usepackage{cite}
\usepackage{amsmath, amsthm}

\usepackage[T1]{fontenc}
\usepackage{inconsolata}

\newtheorem{phase}{Phase}
\newtheorem{factor}{Factor}

\DeclareMathOperator{\deriv}{\mathrm{d}}

\title{Fairness of Congestion-Based Congestion Control: Experimental Evaluation and Analysis}
\author{
	\IEEEauthorblockN{
		Shiyao Ma\IEEEauthorrefmark{1},
		Jingjie Jiang\IEEEauthorrefmark{1}, 
		Wei Wang\IEEEauthorrefmark{1}, 	
		 Bo Li\IEEEauthorrefmark{1}}

		\IEEEauthorrefmark{1}{Department of Computer Science and Engineering, Hong Kong University of Science and Technology} 

}

\begin{document}
\maketitle
\newcommand{\paper}{BBQ\xspace}
\newcommand{\figref}[1]{Fig.~\ref{fig:#1}}

\begin{abstract}
  BBR is a new congestion-based congestion control algorithm proposed by Google.
  A BBR flow sequentially measures the bottleneck bandwidth and round-trip delay of 
  the network pipe, and uses the measured results to govern its sending behavior, maximizing
  the delivery bandwidth while minimizing the delay. However, our deployment in
  geo-distributed cloud servers reveals a severe RTT fairness problem:
  a BBR flow with longer RTT dominates a competing flow with shorter RTT.

  Somewhat surprisingly, our deployment of BBR on the
  Internet and an in-house cluster unearthed a consistent bandwidth disparity among competing flows.
  Long BBR flows are bound to seize bandwidth from short ones.  Intrigued by this unexpected
  behavior, we ask, is the phenomenon intrinsic to BBR? how's the severity? and what's the root
  cause?  To this end, we conduct thorough measurements and develop a theoretical model on bandwidth
  dynamics.  We find, as long as the competing flows are of different RTTs, bandwidth disparities
  will arise.  With an RTT ratio of 10, even flow starvation can happen.  We blame it on BBR's
  connivance at sending an excessive amount of data when probing bandwidth.  Specifically, the amount
  of data is in proportion to RTT, making long RTT flows overwhelming short ones.   Based on this observation, we design a derivative of BBR that achieves guaranteed flow fairness, at the
  meantime without losing any merits.  
  We have implemented our proposed solution in Linux kernel and evaluated it through extensive experiments.
\end{abstract}
\section{Introduction}
\label{sec:intro}

BBR \cite{cardwell2017bbr} (Bottleneck Bandwidth and RTT) emerges as a new congestion-based
TCP congestion control algorithm that, for the first time, converges to
Kleinrock's optimal operating point \cite{kleinrock1979power}, maximizing
delivery rate while minimizing round-trip time (RTT). Unlike
traditional loss- or delay-based congestion control (e.g., \cite{jacobson1988congestion, newreno,
bic, ha2008cubic, vegas, timely}), BBR does not passively react to packet loss or delay. Instead, it takes an initiative stance by
sequentially probing bottleneck bandwidth and minimum round-trip time, and using those
measurements to deliver at full bottleneck bandwidth without creating an excess queue in the pipe. BBR has been added into Linux network stack since mainline 4.9.
According to Google \cite{cardwell2017bbr},
the deployment of BBR has brought about up to $133 \times$ bandwidth
improvement in B4 \cite{jain2013b4} and more than 80\% reduction of the median RTT
in YouTube.



\begin{figure}[t]
	\centering
	\includegraphics{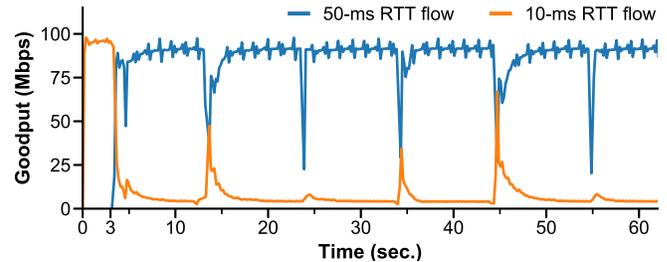}
	\caption{The 50-ms RTT flow grabs most of the bandwidth. Excluding the startup phase, the 50-ms RTT flow has an average goodput of 87.3~Mbps, while the 10-ms RTT flow only has 6.3~Mbps.}
	\label{fig:example}
	\vspace{-.15in}
\end{figure}

Attracted by its promising performance advantages, we deployed BBR in
our geo-distributed cloud servers, yet run into
significant, unexpected RTT fairness issues. Specifically, a flow with longer RTT
always overwhelms those with shorter RTTs. To confirm that such a phenomenon is
not due to the noisy environments in the Internet, we have reproduced the same problem 
in our in-house cluster, where two flows, one with RTT of 10~ms and
the other with RTT of 50~ms, compete on a bottleneck link of 100~Mbps. As shown
in Fig.~\ref{fig:example}, the 10-ms RTT
flow, when running alone in the beginning, delivers at full bandwidth. However,
later when the 50-ms RTT flow joins, the 10-ms RTT flow quickly gives up bandwidth and settles on 6.3~Mbps
most of the time.

%

This outcome is particularly surprising since traditional TCP congestion
control algorithms, be it loss- or delay-based, all favor flows
with shorter RTTs \cite{bic, ha2008cubic, lakshman1997, brown2000, dctcp-sigmetrics}. BBR, on the contrary, has a bias \emph{against} them. Such
a unique bias against flows with shorter RTTs has two serious consequences. 
First, it presents an unpleasant tradeoff between low latency
and high delivery rate, unjustifying the decades of engineering efforts of
bringing down end-to-end latency. For example, finding a route with the minimum RTT
using the OSPF protocol would no longer be desirable, as flows along that route
are easily overwhelmed when competing with others traversing along a suboptimal route
with higher latency. Second, to make things worse,
the advantage of a long RTT flow exposes a loophole which allows a strategic receiver to 
steal bandwidth from competing flows by artificially adding
RTT latency to its inbound traffic.



Concerning about these consequences, 
in this paper, we seek to answer the following three questions:





\textbf{\emph{1) How significant is the bias against short RTT?}}
We performed comprehensive measurement studies in a clean network environment
in an in-house cluster with 20
servers, and made the following three major observations. 

First, BBR flows with shorter RTTs are always unfavored when competing against
those with longer RTTs, irrespective of the bottleneck bandwidth, deployment
of AQM strategies, RTT difference, and the number of competing flows.

Second, a small RTT disparity is sufficient to result in a significant
difference in throughput. In our experiments, a 10-ms RTT flow, when competing with a
15-ms RTT flow, ends up with $< 25\%$ of the bottleneck bandwidth. The larger the
RTT difference is, the more salient the bandwidth disparity would be: 
when competing with a flow with RTT $\ge30$ ms,
the 10-ms RTT flow is squeezed to $< 10\%$ of the bandwidth.

Third, the advantage of a long RTT flow scales. A 50-ms RTT flow occupies $50\%$
of the bandwidth when competing with twenty 10-ms RTT flows. 




\textbf{\emph{2) What is the root cause of the bias?}} Through in-depth analysis of the behaviors of the two competing flows in
\figref{example}, we show that the bias is introduced in two phases. (1) An
excess queue forms on the bottleneck and grows quickly when BBR flows increase inflight to probe for more bandwidth. (2) A long RTT flow floods in a larger volume of excess traffic ($\texttt{inflight} - \texttt{BDP}$) than a short RTT flow, dominating the queue backlog as well as the delivery rate. The short RTT flow measures a lower delivery rate and slows down to match the measurement, making itself more of an underdog in the competition. Worse, the short RTT flow is susceptible to being \texttt{CWND}-bounded, thus unable to probe for more bandwidth.



\textbf{\emph{3) How can we mitigate such a bias?}}
Based on our answer to the previous question, we propose \paper, a simple, yet
effective solution that provides better RTT fairness
without deviating from Kleinrock's optimal operating point. BBQ constantly
detects the presence of an excess queue in the pipe. When a queue forms,
\paper prevents long RTT flows from pouring an overwhelming amount of excess traffic. To do so, \paper enforces a small length of probing period to restrict flows from probing too long.
Conversely, when the queue dissipates, \paper switches to a longer probing period. This allows flows to quickly probe for the available bandwidth, ensuring high link utilization.


We implemented \paper in Linux kernel 4.9.18 and evaluated its performance in our 20-machine cluster through comprehensive experiments. Evaluations show that compared with BBR, \paper significantly improves RTT fairness, increasing the bandwidth share of a short RTT flow by up to 6.1$\times$. 
Such a fairness improvement is achieved without compromising the full delivery bandwidth and low latency.
Moreover, with \paper, the average queueing delay is further reduced by 64.5\%.

\section{BBR: Congestion-Based Congestion Control}
\label{sec:bbr}

In this section, we give a brief introduction on how BBR works and what
benefits it provides. For more details on the implementation, we refer the interested readers to Cardwell et al.~\cite{cardwell2017bbr}.

\subsection{The Optimal Operating Point}

For a TCP connection, there exists one slowest link at a time, known as the \emph{bottleneck}. The bottleneck determines
the connection's maximum delivery rate and is the only place where persistent
queues build up. Ideally, the connection should (1) send at a rate matching
the bandwidth available at the bottleneck (denoted as  \texttt{BtlBw}), and
(2) maintain the amount of data \emph{in flight} that matches exactly one
bandwidth-delay product (BDP), i.e., \texttt{inflight} = \texttt{BtlBw}
$\cdot$ \texttt{RTprop}, where \texttt{RTprop} is the round-trip
propagation time. Kleinrock proved that this operating point maximizes the connection's throughput
(fills the pipe) while minimizing RTT (keeping queues empty), and is optimal
for both individual connections and the network as a whole
\cite{kleinrock1979power}.

However, prevalent TCP congestion control algorithms do not converge to
Kleinrock's optimal operating point. Most of these algorithms use packet loss
as a congestion signal (notably Reno \cite{jacobson1988congestion} and its
successor CUBIC \cite{ha2008cubic}), delivering at full pipe bandwidth at the cost
of \emph{bufferbloat} \cite{bufferbloat}. When the buffer
is deep, which is commonly observed on the last-mile links of today's
Internet, the resulting bufferbloat can easily cause queueing delay of
seconds.

Converging to the optimal operating point has long been a challenging problem,
because \texttt{BtlBw} and \texttt{Rtprop} cannot be measured at the same time. To
measure \texttt{BtlBw}, the pipe must be overfilled, and persistent queues
form; to measure \texttt{RTprop}, on the other hand, all queues must be 
drained empty.





\subsection{BBR's Principles and Benefits}

The recently proposed BBR is a ground-up redesign of congestion control algorithm
\cite{cardwell2017bbr}. BBR addresses the challenge of finding the optimal operating point by
sequentially measuring a connection's maximum delivery rate and minimum RTT.


\vspace{.4em}
\noindent \textbf{Principles.}
In a nutshell, BBR estimates \texttt{BtlBw} as the maximum delivery rate in
recent 10 round trips and \texttt{RTprop} as the minimum RTT
measured in the past 10 seconds. The max-filtered bandwidth (\texttt{MaxBw}) and the min-filtered RTT (\texttt{MinRTT}) precisely model
the pipe.

Based on this model, BBR governs its sending behavior through two control
parameters: pacing rate and congestion window (\texttt{CWND}). BBR cycles
through different pacing rates to probe for more bandwidth (i.e., pacing 25\% faster than
\texttt{MaxBw}), drain off excess queues (i.e., pacing 25\% slower than \texttt{MaxBw}),
and cruise at the current \texttt{MaxBw} to fully utilize bandwidth. BBR also sets \texttt{CWND} as a
small multiple of BDP (2$\times$ by default), so as to bound the volume of
inflight data without creating long, persistent queues.

\vspace{.4em}
\noindent \textbf{Behaviors.}
When a BBR flow connects, it starts by performing an exponential search for the bottleneck bandwidth
by increasing its sending rate by a factor of $2/\boldmath{\ln} 2$ while the delivery rate is growing. Once
\texttt{BtlBw} has been detected, the flow transitions into the \texttt{Drain} mode, clearing up the
excess queue during the search. Then BBR exits this exponential-growth startup and enters the steady
state.

In the steady state, a BBR flow alternates between \texttt{ProbeBw} and 
\texttt{ProbeRTT} mode to dynamically characterize the
pipe's \texttt{BtlBw} and \texttt{RTprop}, based on the recently measured
delivery rates and RTTs. A long-lived BBR flow spends the vast majority of its
time in \texttt{ProbeBW}, pacing at different rates to fully probe and
utilize the pipe's available bandwidth, while maintaining a small, bounded queue. If a flow has
been continuously sending, and has not seen an RTT measurement that matches or
reduces its \texttt{MinRTT}  for a long time (10 seconds by
default), it will briefly enter the \texttt{ProbeRTT} mode to cut the
inflight to a very small value (set \texttt{CWND} to four packets) 
to re-probe the round-trip propagation delay.

\vspace{.4em}
\noindent \textbf{Benefits.} BBR has quickly attracted a wide range of
attention since its publication, due to the following promising benefits:

\begin{itemize}
	\item \emph{Easy to deploy:} BBR is a sender-based congestion control. It requires no modification of switches, nor does it need support for special functionalities, such as ECN \cite{dctcp-sigmetrics}, RDMA \cite{timely} or per-flow AQM \cite{codel}. 
	
	\item \emph{Near-optimal latency:} BBR runs with nearly empty queue most of the time. Furthermore,
	BBR at most uses one BDP's worth of queue by explicitly bounding its inflight.
	
	\item \emph{High throughput:} 
	BBR quickly saturates high capacity links despite the existence of random losses and link errors. In lossy network environments, BBR can significantly improve throughput compared with loss-based congestion control.
\end{itemize}

Encouraged by these promising benefits, we deployed BBR in our
geo-distributed cloud servers to speed up bulk transfer, yet encountered
unexpected bias towards long RTT flows. We next reproduce these problems in a more controlled manner.

\section{Bias towards Long RTT}
\label{sec:measurement}

In this section, we study BBR's RTT fairness performance through several benchmarks in an in-house cluster. Our measurement shows that flows with long RTTs are bound to overwhelm those with shorter RTTs. 
We then discuss the practical concerns about this bias.


\subsection{Methodology}

In order to eliminate the uncontrollable interferences in the wild Internet (e.g., background traffic,
specialized token-bucket policers, AQM deployment, etc.), 
we purposely performed the measurements on an in-house cluster of 20 blade
servers. Each server has a Xeon E5-1410 8-thread 2.8~GHz CPU with 24~GB RAM, and is
interconnected through 1~Gbps NIC. We deployed Linux 4.9.18 with
the latest BBR kernel module across the cluster. The \texttt{sch\_fq} module is
loaded for flow pacing as required by BBR. 

\begin{figure}[tb]
  \centering
  \includegraphics{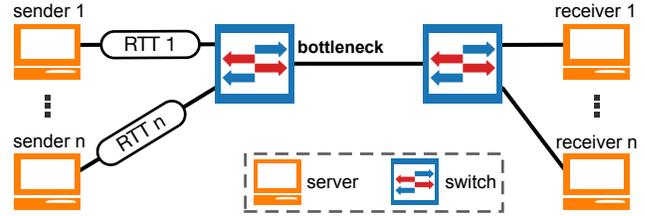}
  \caption{The network topology of our in-house cluster.}
  \label{fig:basic-topo}
  \vspace{-.15in}
\end{figure}

\figref{basic-topo} shows the network topology of our cluster. 
Sender~\textit{n} connects to receiver~\textit{n} with a configurable RTT governed by \texttt{NetEm} \cite{netem}. 
All flows share the same bottleneck link.  A powerful 20-port blade server running on Linux acts as the
bottleneck switch for controllable bandwidth. The switch buffer size per port is by default 2~MB, which is a moderate number
among mainstream products such as Cisco Nexus 3064X, Arista 7050S-64, Juniper QFX3500, etc.

To examine BBR's RTT fairness in different scenarios, we developed a dedicated
framework to coordinate and synchronize the measurement performed on each
server. Specifically, we used \texttt{Netperf}\cite{netperf} to generate
flows, and used the Linux \texttt{tc} tool to control packet delay, bottleneck
bandwidth, and AQM deployment. Our framework has been optimized so that TCP Small Queues\cite{tsq}, a flow control enhancement in modern Linux kernel, will not interfere with 
\texttt{NetEm}\cite{netem} in \texttt{tc}.


\subsection{Micro-benchmark}

We are curious to know if the bias towards long RTT flows observed in
\figref{example} is an intrinsic problem of BBR, or merely an edge case
occurring at some extreme operating parameters. To answer this
question, we ran several micro-benchmarks under different network configurations. Unless
otherwise stated, the RTT of a flow is
configured to be either 10~ms or 50~ms; each flow lasts for 120~seconds for bulk transfer.

\begin{figure}[htb]
  \centering
  \includegraphics{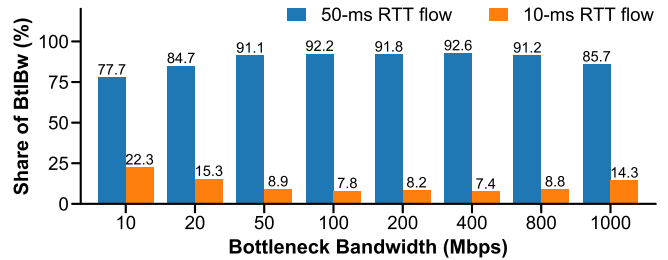}
  \caption{Bandwidth share of two BBR flows competing on a bottleneck link of different
  capacities.}
  \label{fig:ubi-varying-bw}
  \vspace{-.1in}
\end{figure}

\vspace{.4em}
\noindent \textbf{Bottleneck bandwidth.}
We show that BBR's bias towards long RTT persists across a wide range of bottleneck bandwidth.
We gradually increase the bottleneck bandwidth from 10~Mbps to 1~Gbps.
For each bandwidth setting, we initiated the same two flows as in the previous experiments,
and measured their bandwidth share in the steady state. 
\figref{ubi-varying-bw} depicts the results. While the bias against the short RTT flow is 
alleviated on low bandwidth bottleneck, the 10-ms RTT flow remains far below its fair share.

\begin{figure}[htb]
  \centering
  \includegraphics{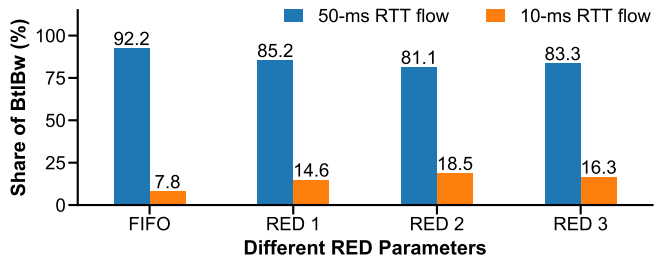}
  \caption{Bandwidth share of two BBR flows competing on a bottleneck with different
  	RED parameters (details in Table~I).}
  \label{fig:ubi-varying-aqm}
  \vspace{-.15in}
\end{figure}

\noindent \textbf{Deployment of AQM.} 
Modern switches employ AQM (active queue management) schemes, notably RED
\cite{floyd1993random} (random early detection), to alleviate the TCP collapse
\cite{allman2009tcp} problem.  \figref{ubi-varying-aqm} shows that raising the drop probability or reducing the \texttt{max-threshold} (details in Table~\ref{table:aqm}) helps the short RTT flow to regain slightly more bandwidth share. However, if one tries to follow this trend and tune the parameters, an enormous amount of retransmissions will be incurred.

\begin{table}[htb]
  \footnotesize
  \centering
  \caption{RED Parameters and Retransmissions}
\begin{tabular}{ | r | c | c | c || c | c |} 
  \hline
  \textbf{Scheme} & \textbf{Max} & \textbf{Min} & \textbf{Prob.} & \textbf{Retrans-50} & \textbf{Retrans-10} \\
  \hline
  RED~1 & 0.50~MB & 0.17~MB & 2\%  & 10178~pkts & 2871~pkts \\
  \hline
  RED~2 & 0.50~MB & 0.17~MB & 10\% & 20583~pkts & 6014~pkts \\
  \hline
  RED~3 & 0.33~MB & 0.17~MB & 2\%  & 17193~pkts & 5301~pkts \\
  \hline 
\end{tabular}
\label{table:aqm}
\vspace{-.1in}
\end{table}

\begin{figure}[htb]
  \centering
  \includegraphics{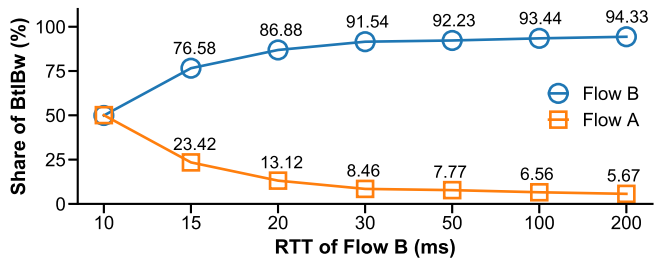}
  \caption{The bandwidth disparity of two BBR flows becomes more salient as their RTT difference increases.}
  \label{fig:ubi-varying-rtt}
  \vspace{-.1in}
\end{figure}

\vspace{.4em}
\noindent \textbf{Disparity of RTT.}
In order to understand how bandwidth share changes with an increasing RTT disparity, we consider the throughput of two competing BBR flows with short
(flow A) and long (flow B) RTTs on a bottleneck link of 100 Mbps.  Flow A has a
fixed RTT of 10~ms; flow B has varying RTTs, ranging from 10~ms to 200~ms. This
range of RTTs captures most LAN and long-haul connections. 
\figref{ubi-varying-rtt}  shows the measured throughput of the two flows. When the two flows are of
the same RTT, fairness is not a concern
\cite{cardwell2017bbr}.  However, as the disparity of RTT increases,
the bias towards long RTT becomes more pronounced. Specifically, it does not
require a large RTT disparity to observe a salient throughput difference: The 15-ms RTT
flow dominates the 10-ms RTT flow with 3$\times$ throughput.\footnote{Google has
recently acknowledged the advantage of long RTT flows, but claimed that
the problem is not of a significant concern of BBR \cite{cardwell2016slides}. We
suspect that this is because Google's experiment was performed in a 
low-bandwidth environment (10 Mbps), where RTT bias is less severe (cf. \figref{ubi-varying-bw}).}  This result is particularly
disturbing. It suggests that  a strategic receiver can steal bandwidth by
artificially inflating its RTT. 
We shall discuss this point in detail in the next subsection.

\begin{figure}[t]
  \centering
  \includegraphics{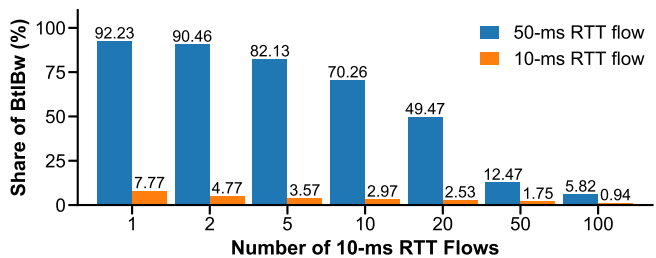}
  \caption{Per-flow bandwidth share of flows with short and long RTTs. 
  A 50-ms RTT flow competes with a varying number of 10-ms RTT flows.}
  \label{fig:ubi-varying-n}
  \vspace{-.15in}
\end{figure}

\vspace{.4em}
\noindent \textbf{Number of competing flows.} 
The number of competing flows is another critical factor affecting bandwidth share. To quantify its impact, we initiated a 50-ms RTT
flow along with a varying number of 10-ms RTT flows. As shown in 
\figref{ubi-varying-n}, the advantage of the long RTT flow diminishes quickly as the number of
competing short RTT flows increases. This is because short RTT flows, even
unfavored, 
can always grab some bandwidth in the end. Therefore, as their number surges, the long RTT flow would end up
with less advantage. Nevertheless, the long RTT flow remains in favor with much
more bandwidth than it deserves even when it is outnumbered by the 10-ms RTT competitors.

\vspace{.4em}
\noindent \textbf{Summary of findings.} Our measurement confirms that 
the RTT fairness problem is intrinsic to BBR.
Flows with long RTTs always have the upper hand. The longer the RTT is, 
the more bandwidth it will get.


\subsection{Practical Concerns}
\label{sec:concern}
BBR's bias towards long RTT flows is in stark contrast to the conventional
wisdom of TCP congestion control. Traditional loss- and
delay-based congestion control algorithms, such as Reno
\cite{jacobson1988congestion}, CUBIC \cite{ha2008cubic}, DCTCP \cite{dctcp},
and Vegas \cite{vegas}, all favor flows with short RTT \cite{lakshman1997,
brown2000, bic, dctcp-sigmetrics}. That BBR has a completely opposite
bias \emph{against} them raises two serious, practical concerns.

\vspace{.4em}
\noindent \textbf{Unpleasant tradeoff.}
First, it enforces an unpleasant tradeoff between low latency and high
delivery rate, which makes no sense in today's Internet. The networking
community has tried for decades to reduce end-to-end latency.
However, in the presence of the latency-bandwidth tradeoff, all those previous
efforts will be unjustified. For instance, finding a route with the minimum
RTT using routing algorithms such as OSPF and IS-IS could turn out undesirable, simply because
flows along the shortest path are easily overwhelmed by others traversing long-haul. 


\vspace{.4em}
\noindent \textbf{Latency-cheating.}
Second, BBR's bias towards long RTTs can be easily manipulated by strategic
receivers, who can steal bandwidth by artificially inflating its RTT
(e.g., delaying inbound traffic). Because BBR is a sender-based congestion
control, it would be very hard, if not impossible, for the sender to tell if
the probed RTT has been manipulated by receivers. 

The consequence of such ``latency-cheating'' can be more complicated. It does
not appear to exist an equilibrium where all receivers are content about their
bandwidth share. This can cause them to continuously game the network
and run into a worse outcome.

\begin{figure}[t]
  \centering
  \includegraphics{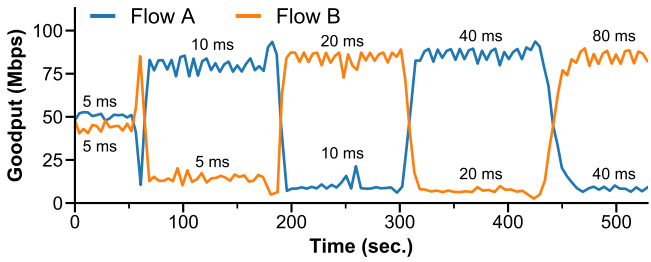}
  \caption{Two flows with 5-ms true RTT alternately cheat on a 100-Mbps link. Each time a flow
  cheats, it inflates its RTT (annotated in the figure) to $2\times$ of the other. The two flows race to the bottom with RTT growing exponentially.}
  \label{fig:strategic}
  \vspace{-.15in}
\end{figure}

To illustrate this ``race-to-the-bottom'' game, we initiated two competing BBR
flows with the true RTT of 5~ms. The two flows alternately cheat.  Each time
a flow cheats, it delays the inbound traffic and inflates its RTT to $2\times$ of
the other. \figref{strategic} shows the measured goodput of the two cheaters
over time. Because a flow can always delay its traffic to steal more
bandwidth from the other, the two keep doing so, causing the RTT
growing exponentially.

We stress that the two concerns above are unique to BBR due to its
congestion-based congestion control. The presence of these concerns clouds the performance
advantages of this promising new algorithm. As a first step to address this problem, 
we next analyze why BBR favors flows with long RTTs.


\section{The Anatomy of Bias}
\label{sec:analysis}

In this section, we analyze the root cause of BBR's bias against short RTT by
diving deep into the behaviors of two competing flows in \figref{example}. We
generalize our findings and show that the bias is developed through two phases.

\subsection{Deep Dive}

At a first glimpse, BBR's bias against short RTT flows appears counter-%
intuitive. BBR flows probe for more bandwidth once in every eight round trips \cite{cardwell2017bbr}. When additional bandwidth becomes available, flows with shorter RTTs
would respond to this environmental change more quickly, and hence have the upper hand to claim more available bandwidth 
in advance.

\begin{figure}[tb]
	\centering	\includegraphics{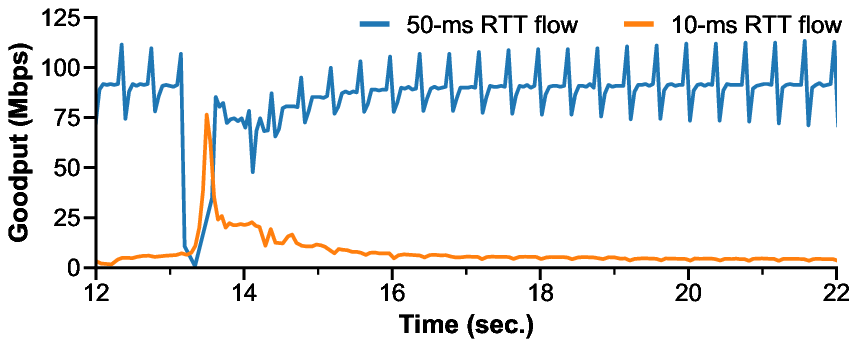}
	\caption{Goodput of the two BBR flows in \figref{example} from 12 to 22 sec.}
	\label{fig:micro-goodput}
\end{figure}

To understand why this first-mover advantage fails to sustain, we first refer back to \figref{example} and focus on the steady-state goodput of the two flows. \figref{micro-goodput} provides a zoom-in view to illustrate more details. For the 50-ms RTT flow, when its \texttt{MinRTT} expires,  it transitions into \texttt{ProbeRTT}: in order to learn the true \texttt{RTprop}, it reduces the inflight to four packets to drain the excess queue. As the queue dips and the bandwidth occupied by the 50-ms RTT flow yields, the 10-ms RTT flow detects a new \texttt{MaxBw} and reclaims all the available bandwidth.
However, this situation does not last long.
When the 50-ms RTT flow returns to \texttt{ProbeBw}, the 10-ms RTT flow is quickly overwhelmed.



\begin{figure}	
	\includegraphics{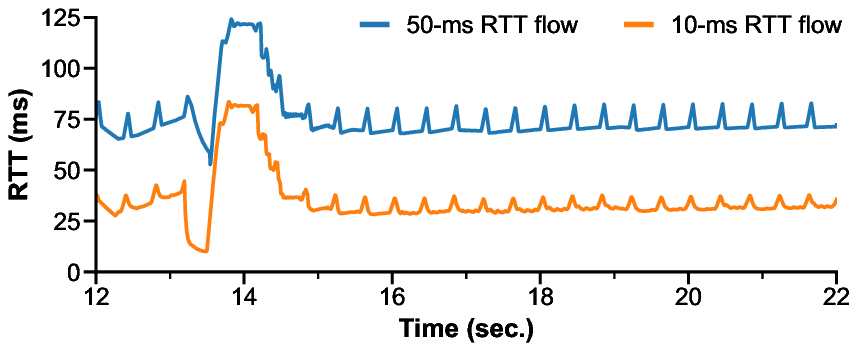}
	\caption{RTT of the two BBR flows in \figref{example} from 12 to 22 sec.}
	\label{fig:micro-rtt}
	\vspace{-.15in}
\end{figure}

Why does the short RTT flow yield the acquired bandwidth to the long RTT flow so
 easily? Our \textbf{first finding} attributes this outcome to \emph{the persistent queue
developed on the bottleneck.}
Note that when the long RTT flow returns to \texttt{ProbeBw}, it resumes
with its previous \texttt{MaxBw} (close to \texttt{BtlBw}). Because the short RTT flow has detected
a much higher \texttt{MaxBw} when the long RTT flow was in \texttt{ProbeRTT}, the aggregated \texttt{MaxBw} of the two flows exceeds the bottleneck capacity, and thus an excess queue develops. \figref{micro-rtt} confirms this theory by tracking
the RTT changes of the two flows over time. After the 50-ms RTT flow exits
\texttt{ProbeRTT}, the RTTs of both flows explode until the surging inflight is bounded
by \texttt{CWND}---a clear sign of a rapid-growing queue 
forming on the bottleneck. 

\begin{figure}[htb]
	\centering
	\includegraphics{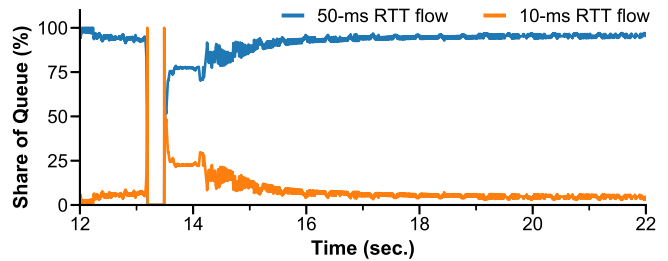}
	\caption{Queue share of the two BBR flows in \figref{example} from 12 to 22 sec.}
	\label{fig:micro-queue}
\end{figure}

Interestingly, the development of the excess queue has drastically different impacts on the
goodput of the two flows. Referring back to \figref{micro-goodput}, for the
short RTT flow, its goodput falls off the cliff as soon as the queue forms,
whereas the goodput of the long RTT flow is decreased by only a small amount.
Why does this happen? Queueing theory tells us that in
the presence of a persistent queue, the bandwidth share of competing flows is
determined by their backlog. This motivated us to measure the instantaneous
queue backlog of the two flows by instrumenting the kernel functions on the bottleneck using
\texttt{SystemTap} \cite{systemtap}.  We see in \figref{micro-queue} that the 10-ms RTT flow quickly occupies all
queue slots when its competitor enters \texttt{ProbeRTT}, meaning that
it has probed for more bandwidth. The 10-ms RTT flow then drains off queues and tries to
cruise at the new bandwidth. However, when the 50-ms RTT flow returns to \texttt{ProbeBw},
the 10-ms RTT flow quickly recedes, and never finds a chance to come back.


We now reach our \textbf{second finding}: \emph{as the queue develops, the
backlog of the flow with shorter RTT is diminishing, so is the bandwidth
share.} To explain this outcome, let us replay what has happened since the return
of the 50-ms RTT flow from \texttt{ProbeRTT}. Both flows enter
\texttt{ProbeBw}, probing for more bandwidth by periodically pouring slightly
more inflight than their own estimated BDP (1.25 BDP) into the pipe. In the
beginning, because both flows detect roughly the same
\texttt{MaxBw} ($\approx$\texttt{BtlBw}), they send at the same rate, and
the \texttt{inflight}$-$\texttt{BDP} excess (0.25 BDP) is proportional to the
flow's \texttt{MinRTT}. Since the 50-ms RTT flow has a larger \texttt{MinRTT}, it
deposits much more packets into the queue for a longer period than the 10-ms
RTT flow. The increased queue share allows it to operate at a higher delivery rate than its 
competitor. Unable to measure a
higher bandwidth, the 10-ms RTT flow quickly expires its \texttt{MaxBw} 
and replaces it with a smaller value. This, in turn, forces the 10-ms flow
to send less data, which results in an even smaller queue share
and a lower delivery rate. A positive feedback loop establishes and drives the 10-ms RTT
flow to the bottom.



\begin{figure}[htb]
	\centering
	\includegraphics{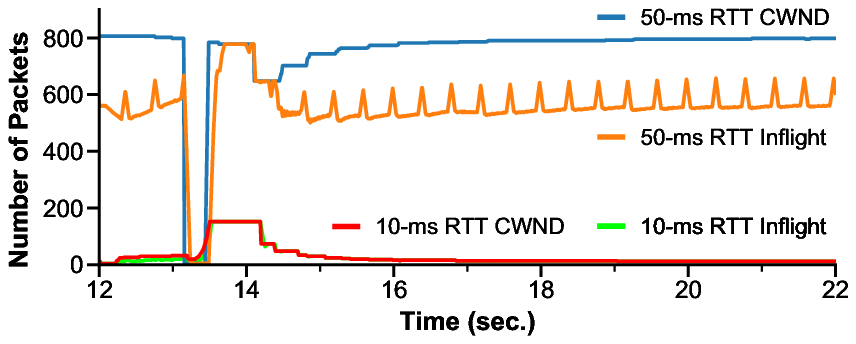}
	\caption{\texttt{CWND} and inflight of the two BBR flows in \figref{example} from 12 to 22 sec.}
	\label{fig:micro-cwnd}
	\vspace{-.1in}
\end{figure}

To make things worse, when the large flow returns to \texttt{ProbeBw}, the
excess queue quickly builds up, and if the queue grows too fast, the inflight of
the short RTT flow would be \texttt{CWND}-bounded
(\texttt{inflight}=\texttt{CWND}) due to the long queueing delay.
Meaning, the flow's sending rate is ACK clocked \cite{jacobson1988congestion} and cannot be further increased. Being
\texttt{CWND}-bounded essentially refrains the short RTT flow from injecting
more packets into the pipe, speeding up the race
to the bottom.  \figref{micro-cwnd} confirms this problem. 
The 10-ms RTT flow has its inflight \texttt{CWND}-bounded, whereas the 50-ms RTT flow can still inflate inflight (shown as the
small spikes) to probe for more bandwidth.



\subsection{The Development of Bias in Two Phases}

Our findings based on the experiment of \figref{example} highlight that the bias
against short RTT is developed in two phases: (1) a persistent queue forms on
the bottleneck, and (2) the backlog of the short RTT flow diminishes. We now
generalize our findings to two competing flows with different RTTs and discuss
the root cause behind. We limit our discussions to the period during which
both flows 
stay in \texttt{ProbeBw}.



\begin{phase}
When the two flows enter \texttt{ProbeBw}, a persistent queue forms on the bottleneck.
The queue keeps growing until one flow is \texttt{CWND}-bounded.
\end{phase}

We explain how this happens through a step-by-step analysis of the behaviors of two
competing flows A and B.

\emph{\textbf{Step 1.}} To probe for more bandwidth, one flow, say flow-A, paces
slightly faster than its \texttt{BtlBw} estimate (\texttt{MaxBw}), inflating
its inflight to 1.25 BDP. An excess queue hence forms on the bottleneck, at
least in one round trip.

\emph{\textbf{Step 2.}} The increased inflight raises the queue share of
flow-A,\footnote{Flow-B is likely not probing at the same time and maintains
its current backlog in the queue, if any.} which, in turn, increases its
delivery rate. Flow-A measures a new maximum bandwidth \texttt{MaxBw}.


\emph{\textbf{Step 3.1.}} Flow-A drains the excess queue after probing. It uses the
newly updated \texttt{MaxBw} and reduces its inflight to a BDP higher than
the previous estimate. Meaning, the excess queue due to probing 
cannot be drained empty.

\emph{\textbf{Step 3.2.}} Flow-B maintains its sending rate and inflight as long
as its \texttt{MaxBw} has not expired. 
The queue keeps growing until one flow expires its \texttt{MaxBw}, after which
the aggregated sending rate falls to match the bottleneck capacity, and the queue
sustains.

Repeating the entire process, we see that the queue continues growing after
each probe-and-drain, until one flow is \texttt{CWND}-bounded and stops
probing.

\begin{phase}
As the queue develops, the backlog of the flow with shorter RTT diminishes, and
the flow is overwhelmed.
\end{phase}

Once a persistent queue forms on the bottleneck, the throughput of a flow is
determined by its queue share. Two factors come into play, respectively 
corresponding to BBR's two control parameters, pacing rate and \texttt{CWND}.

\begin{factor}[Positive feedback loop]
  The flow with shorter RTT tends to contribute less queue backlog due to a
  smaller BDP estimate. This reduces its queue share and triggers a positive 
  feedback loop with even less share.
\end{factor}

We give an intuitive explanation. A BBR flow periodically paces faster than the current
\texttt{BtlBw} estimate to probe for more bandwidth, depositing 0.25 BDP worth
of excess traffic to the queue. Intuitively, a flow with shorter RTT has a
smaller BDP estimate than its competitor, and injects much less excess traffic
into the pipe. This drives its queue share down, followed by a lower delivery
rate. The persistently low delivery rate tricks the flow to lower the
\texttt{BtlBw} estimate and sends at a lower rate, which further reduces its
queue share, triggering a positive feedback loop.


\begin{factor}[\texttt{CWND}-bounded inflight]
	To make things worse, as the queue develops, the flow with shorter RTT tends to be \texttt{CWND}-bounded, and is unable to probe for more bandwidth.
	\label{thm:cwnd}
\end{factor}

We explain how this happens in two steps.

\emph{\textbf{Step 1.}} The flow with shorter RTT enters the \texttt{CWND}-bounded
mode before its competitor.
According to BBR \cite{cardwell2017bbr}, a flow is \texttt{CWND}-bounded if and only if the queueing delay is greater than the \texttt{minRTT}, i.e.,
 \vspace{-.05in}
\begin{equation}
  \label{eq:cwnd-bounded}
  \texttt{inflight} = \texttt{Bw}\cdot\texttt{RTT} > 2\cdot \texttt{MaxBw}\cdot\texttt{MinRTT}.
\end{equation}
Because the two flows share the same bottleneck queue, 
as the queue develops, the queueing delay keeps increasing and will
exceed a smaller \texttt{MinRTT} first. Meaning,
a flow with shorter RTT is \texttt{CWND}-bounded first.

\emph{\textbf{Step 2.}} Once entering the \texttt{CWND}-bounded mode, the flow
will stay in it throughout \texttt{ProbeBw}.
This is because as the queue develops, the \texttt{CWND}-bounded flow measures increasing RTT. It hence
has no chance to find a way out as
its queueing delay is always greater than its \texttt{MinRTT}.

Now that the short RTT flow is \texttt{CWND}-bounded, it loses control of its
sending rate which is ACK clocked \cite{jacobson1988congestion} (matching the delivery rate).
Since then, the queue stops growing, and the queueing delay remains unchanged.
This suggests that the long RTT flow will never be \texttt{CWND}-bounded
(its \texttt{MinRTT} is always greater than the queueing delay).






\section{Improving RTT Fairness using BBQ}
\label{sec:design}


In this section, we close the gap between flows with different RTTs using a
BBR improvement algorithm. Our design goal  is to \emph{provide better RTT
fairness than BBR without deviating from Kleinrock's optimal operating point},
i.e., maximizing delivery rate while minimizing latency.

Our previous discussions reveal that the source of bias against
short RTT flows originates from the rapid growth of a persistent queue when
flows are probing for more bandwidth. Therefore, preventing a queue from building up too
fast in \texttt{ProbeBw} is the key to achieving better RTT fairness. 

As a first-cut solution, we consider a simple approach that completely gets rid of
the queue once it forms after the probing, before it causes any damage.

\subsection{Too Late to Drain after Probing}


BBR already drains queues after probing. However, this is performed only in best
efforts, and the queue may not be drained empty. Specifically, in the
\texttt{ProbeBw} mode,  BBR has a drain period following the probing period, during
which a flow paces slower than its \texttt{MaxBw}, keeping the inflight to 0.75 BDP
estimate (setting pacing gain to 0.75). The drain period ends when the inflight falls
below one BDP (no excess queue), \emph{or} the period has spanned one \texttt{MinRTT}, whichever comes
first \cite{bbr-commit}. If it is the latter that comes first, the
\texttt{$\mbox{inflight} - \mbox{BDP}$} excess is not fully cleared up.

Instead of trying best-effort, our implementation forces each flow to drain
inflight to \emph{exactly one} up-to-date BDP estimate. This way, the flow
leaves no backlog in the queue, so that long RTT flows cannot squeeze out
short ones.

\begin{figure}[htb]
	\centering
	\includegraphics{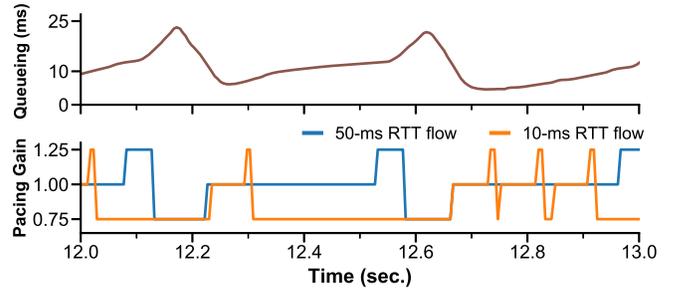}
	\caption{The queueing delay and gain cycling of a 50-ms RTT flow and a 10-ms RTT flow under the first-cut solution.}
	\label{fig:sol-naive}
	\vspace{-.1in}
\end{figure}

Contrary to our expectation, such a drain-after-probing approach is of little
help in improving RTT fairness. Applying this approach to the previous experiment in
\figref{example}, we found that the short RTT flow ends up with even lower bandwidth
share than that using BBR. Curious about why this happened, we dig into the
pacing cycles of the two flows. As shown in \figref{sol-naive}, 
the 10-ms RTT flow spends the majority of its time in the drain periods, and when it
proceeds to cruising (setting pacing gain to 1), its \texttt{MaxBw} has long expired
and is replaced with a smaller estimate. This forces
the short RTT flow to send at a lower rate, hence triggering a
positive feedback loop.


Why is the 10-ms RTT flow trapped in the draining period for so long? In general,
with a pacing gain of 0.75, each flow can at best drain inflight to $0.75
\cdot \texttt{MaxBw} \cdot (\texttt{MinRTT + Queueing})$. Therefore, if the
queueing delay is persistently larger than $\frac{1}{3}\texttt{MinRTT}$, the
flow is unable to drain its inflight to one BDP. As illustrated in \figref
{sol-naive}, while the 10-ms RTT flow keeps trying to drain its inflight, the
queueing delay is still increasing due to the long RTT flow probing at the same
time. The 10-ms RTT flow is then trapped in the drain periods until its competitor
finishes probing and starts to drain.


To summarize, once a long, persistent queue has formed, it will be too late to get rid of it.




\subsection{\paper}


The failure of the drain-after-probing approach prompts us that the
algorithm must react to the queue development early on. Specifically, the
algorithm should constantly detect if an excess queue is forming on the
bottleneck. When it happens, the algorithm quickly intervenes,
refraining long RTT flows from pouring too much excess traffic into the pipe,
so as to protect short RTT flows from being squeezed out.

Our solution, which we call \paper, follows exactly this intuition. We start
to focus on how BBQ regulates excess traffic poured into the pipe. We
then discuss how BBQ detects the presence of a persistent queue.


\vspace{.4em}
\noindent \textbf{Regulating excess traffic.}
In order to probe for more bandwidth, a BBR flow periodically pours excess
traffic into the pipe when it comes to the probing period. A probing period
spans one \texttt{MinRTT}, during which the flow paces 25\% faster (sets pacing
gain 1.25) than the measured \texttt{MaxBw}.  Because a long RTT flow
probes for a
longer time (cf. \figref{sol-naive}), it pours more excess traffic into the pipe and thus dominates short RTT flows.

To eliminate this advantage for long RTT flows, we impose a 
\emph{cap} to the span of a probing period. Instead of probing for
\texttt{MinRTT} time, our solution, BBQ, employs the following length for the probing period:
  \vspace{-.05in}
\[
  \rho = \min \{\texttt{MinRTT}, \alpha \},
  \vspace{-.04in}
\]
where $\alpha$ is a parameter that caps the probing period.
Imposing a cap to the probing period enables BBQ to regulate excess 
traffic, improving RTT fairness from two perspectives.

First, a long RTT flow with $\texttt{MinRTT}>\alpha$ now probes for at most
$\alpha$ time and will not flood in an overwhelming amount of excess inflight
to edge out the coexisting short RTT flows. With a small $\alpha$ (e.g., 3 ms),
flows probe for the same period of time, making nearly equal contributions
to the bottleneck queue with fair bandwidth share 
(details in Sec.~\ref{sec:eval}).

Second, regulating excess traffic prevents a queue from growing too
fast or too long. The reduced queueing delay lowers the chance of having
a short RTT flow \texttt{CWND}-bounded.

We stress that the cap is imposed to a probing period only, while the length of
the other periods (draining and cruising) remains unchanged. Similar to BBR,
BBQ never over-drains inflight, and stops draining at one BDP 
or when the period has span one \texttt{MinRTT}, whichever comes first. 
BBQ hence delivers at the maximum bandwidth with better RTT
fairness than BBR.




Regulating the excess traffic by capping the probing period is not without its
downside. In general, it slows down the probing for more bandwidth. In case
that more bandwidth becomes available (e.g., a route update or the departure
of a competing flow), flows cannot quickly ramp up and may take
a long time to saturate the bottleneck. 

BBQ addresses this problem by judiciously capping the probing period only when
the bottleneck is fully utilized with a persistent queue. In the absence of a
persistent queue, BBQ simply reduces to BBR, allowing flows to quickly probe
for more bandwidth. This requires BBQ to constantly detect the presence of a
non-zero queue.

\vspace{.4em}
\noindent\textbf{Queue detection.}
Existing works \cite{vegas, timely} interpret a negative delay gradient
measured through RTT signals ($\frac{\deriv q(t)}{\deriv t} = \frac{\deriv
\texttt{RTT}}{\deriv t}$) as an indicator for the decreasing contention on the
bottleneck, provided that a non-zero queue exists in the network. However,
RTT gradient cannot be used as an evidence for the presence/absence of
a persistent queue here because
the pacing gain cycling in BBR incurs frequent queue size changes. 
In this regard, we turn to a direct RTT measurement instead. Specifically,
\paper considers the pipe underutilized if the RTT measurement drops below
a threshold: 
\[
  \texttt{RTT} < (1 + \beta) \texttt{MinRTT},
\]
where $\beta$ is a slack factor used
in account of inaccurate measurement or unstable link states.
BBQ reduces to BBR when the pipe is underutilized.

\vspace{.4em}
\noindent \textbf{Summary of the algorithm.} 
\paper flows keep detecting the presence of a persistent queue
through RTT measurements. When the pipe is full, \paper flows probe for
approximately the same period of time $\alpha$, so as to have an equal share
of the queue backlog. When the pipe is underutilized, \paper simply reduces to
BBR. This way, flows can quickly probe for the available bandwidth and utilize
it as fast as possible.



\subsection{Guidelines for Choosing Parameters}
\label{sec:guidelines}
\noindent\textbf{Cap for probing period.}
Having a large cap $\alpha$ for a probing period is less beneficial
to improving RTT fairness. For example, if we choose $\alpha=15$~ms for a
10-ms RTT flow coexisting with a 50-ms RTT flow, the former can at most probe
for 10~ms, whereas the latter can probe for 15~ms. The 50-ms RTT flow still
has the upper hand, though less advantageous compared with BBR. In general, to optimize
RTT fairness, it is desirable to have
$\alpha$ smaller than the typical propagation delay of Internet connections. 
On the other hand, having a too small $\alpha$ results in
a slow convergence to the stable bandwidth share. We recommend
$\alpha=3$~ms and will show in Sec.~\ref{sec:parameter} that such a choice fits
most LAN and long-haul connections with propagation delays ranging from 5~ms
to 300~ms.

\vspace{.4em}
\noindent\textbf{Slack factor.} The slack factor $\beta$ must be small enough to avoid false detection of pipe under-utilization, but meanwhile not too small so that high link utilization is ensured. For wired networks, we recommend $\beta  \in [0.5\%, 1\%]$.

\subsection{Limitations}


Similar to BBR, BBQ employs a constant \texttt{CWND} gain to bound the inflight of each
flow to 2 BDP. A short RTT flow hence has a smaller \texttt{CWND} than a long RTT flow, provided
that both flows have the same estimation of the bottleneck bandwidth. This suggests that a short RTT flow will
be \texttt{CWND}-bounded earlier, and will gradually yield bandwidth to the long RTT flow. 
Can we address this problem by raising the \texttt{CWND} bound to a larger multiple of BDP? 
In our experiments, by setting \texttt{CWND} to 4 BDP, we did observe a higher bandwidth share 
for short RTT flows, yet at a price of a significantly longer queueing delay. 
For this reason, \paper chooses to prioritize low latency by retaining \texttt{CWND=2 BDP}. 
How to adaptively adjust the \texttt{CWND} gain to navigate the tradeoff between RTT fairness and queueing delay is left for future work.

\section{Evaluation}
\label{sec:eval}
\begin{figure}[t]
	\centering
	\includegraphics{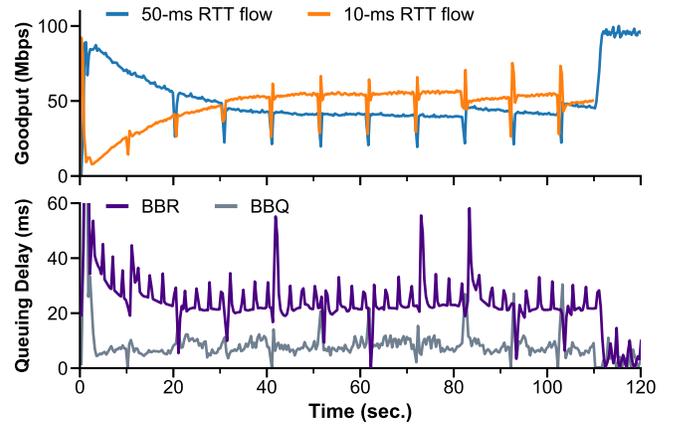}
	\caption{The 10-ms RTT flow has a goodput of 51.4~Mbps in steady state, and the 50-ms RTT flow has 42.5~Mbps. After the 10-ms RTT flow leaves, the 50-ms RTT flow promptly saturates the link. Besides, BBQ flows experience a smaller queueing delay compared with BBR. }
	\label{fig:bbq-overall}
\end{figure}

We have implemented \paper in Linux kernel 4.9.18 and evaluated its
performance in the same cluster environment as described in
Sec.~\ref{sec:measurement}. Highlights of our evaluations are summarized as
follows.


\begin{itemize}
\item \paper delivers at full bandwidth with near-optimal latency. Compared
with BBR, \paper reduces the queueing delay by 64.5\% when a 10-ms RTT flow
is competing with a 50-ms RTT flow (Sec.~\ref{sec:efficiency}).

\item \paper consistently outperforms BBR with better RTT fairness, improving
the bandwidth share of a short flow by up to $6.1\times$ (Sec.~\ref{sec:fairness}).

\item \paper's performance advantage extends to a wide range of RTTs, 
from 5~ms to 300~ms (Sec.~\ref{sec:parameter}).
\end{itemize}

\subsection{Micro-benchmark}
\label{sec:efficiency}

We demonstrate the advantage of \paper through a micro-benchmark in  \figref
{bbq-overall}, where two flows, one with 10-ms RTT and the other with 50-ms
RTT, compete on a bottleneck link of 100~Mbps.  In the startup mode, while
the 10-ms RTT flow occupies the full bandwidth first, it quickly yields to the
50-ms RTT flow, as the latter probes more aggressively during the startup.
Later in the steady state, both flows enters \texttt{ProbeBw}. The short flow
comes back and settles on 51.4 Mbps after the synchronization of
\texttt{ProbeRTT} at 31~s, while the 50-ms RTT flow stabilizes at around 42.5
Mbps. When the 10-ms RTT flow departs at 110~s, BBQ instantly detects this and
reduces to BBR. The 50-ms RTT flow quickly ramps up and takes the remaining
bandwidth in just 1.7~seconds.

In addition to the significantly improved RTT fairness and high throughput,
\paper outperforms BBR with a lower delay. As shown in the lower graph of \figref
{bbq-overall}, \paper reduces the average queueing delay to 8.3~ms, as opposed
to 23.4~ms when using BBR, a $65\%$ reduction on average.

\subsection{Impact of Network Environment}
\label{sec:fairness}

For comparison purpose, we evaluate \paper's RTT fairness performance using
the same benchmarks as in Sec.~\ref{sec:measurement}. Unless otherwise stated,
we let a 10-ms RTT flow compete with a 50-ms RTT flow on a bottleneck link  of
100~Mbps. We measure RTT fairness of a due algorithm (\paper and BBR) using
the bandwidth share received by the short RTT flow. 


\begin{figure}[htb]
	\centering
	\includegraphics{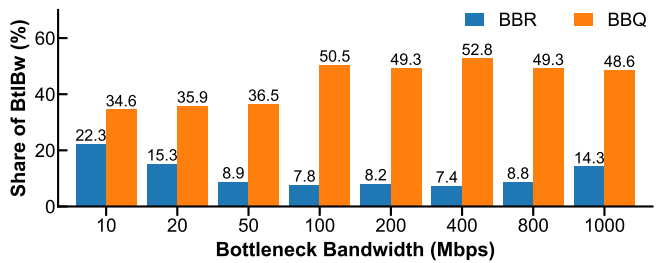}
	\caption{\paper improves the bandwidth share of the 10-ms RTT flow by 55.2\% to 6.14$\times$ compared with BBR under different bottleneck capacities.  }
	\label{fig:sol-varying-bw}
\end{figure}

\vspace{.4em}
\noindent \textbf{Bottleneck bandwidth.}
\figref{sol-varying-bw} compares the bandwidth share a short RTT flow receives
using BBQ and BBR with bottleneck bandwidth spanning two orders of magnitude.
In all cases, BBQ outperforms BBR with better RTT fairness, improving the bandwidth
share of the 10-ms RTT flow by at least $55.2\%$ (10 Mbps) and by up to
$6.1\times$ (400 Mbps). BBQ
provides near-optimal RTT fairness for high-bandwidth link ($\ge 100$~Mbps), where
the two flows equally share the bottleneck.




\begin{figure}[tb]
  \centering
  \includegraphics{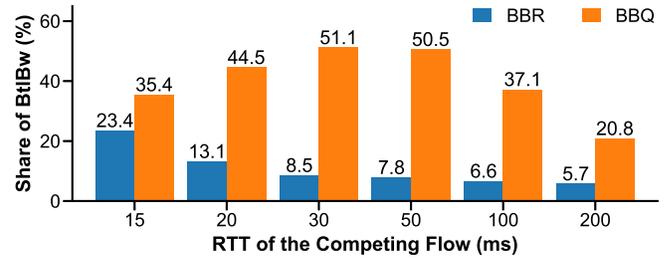}
  \caption{\paper improves the bandwidth share of the 10-ms RTT flow by up to 4.6$\times$ when competing with a flow with varying RTTs. }
  \label{fig:sol-varying-rtt}
\end{figure}

\vspace{.4em}
\noindent \textbf{Disparity of RTT.}
In order to confirm that \paper's improvement in fairness extends to 
different levels of RTT disparity, we let a 10-ms RTT flow compete with a flow
with varying RTTs, ranging from 10~ms to 200~ms. 
\figref{sol-varying-rtt} compares the bandwidth share of the 10-ms RTT flow 
using \paper and BBR. We see that for BBR flows, a larger RTT difference widens their throughput disparity. In comparison,
\paper successfully constrains this trend. Specifically, even when competing with a flow
with $10\times$ RTT (100ms), the 10-ms RTT flow can still retain 37.1\% of the bottleneck bandwidth---a $4.62\times$ fairness improvement over BBR.

\begin{figure}[tb]
	\centering
	\includegraphics{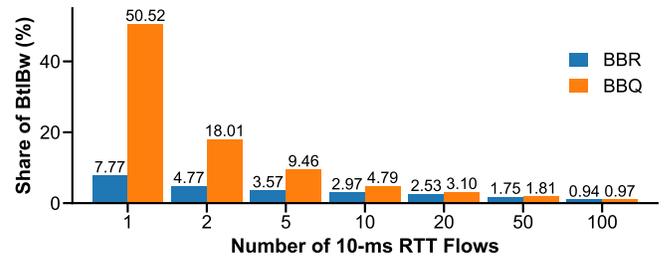}
	\caption{The per-flow bandwidth share of each 10-ms RTT flow with different number of 10-ms RTT flows and one 50-ms RTT flow. }
	\label{fig:sol-varying-n}

\end{figure}

\vspace{.4em}
\noindent \textbf{Number of competing flows.}
We next evaluate if BBQ's improvement in fairness scales to more flows. We let
a varying number of 10-ms RTT flows compete with one 50-ms RTT flow.
\figref{sol-varying-n} compares the bandwidth share of each 10-ms RTT flows
using \paper and BBR. As expected, in all cases, each 10-ms RTT flow improves
its bandwidth share using \paper. However, as their number increases, the
benefits provided by \paper become less salient, for two reasons.
On one hand, the advantage of the 50-ms RTT flow diminishes when it is outnumbered by the 10-ms RTT competitors. On the other hand, with a growing number of competing flows, 
the queueing delay surges, and the 10-ms RTT flow are more likely \texttt{CWND}-bounded.

\begin{figure}[t]
	\centering
	\includegraphics{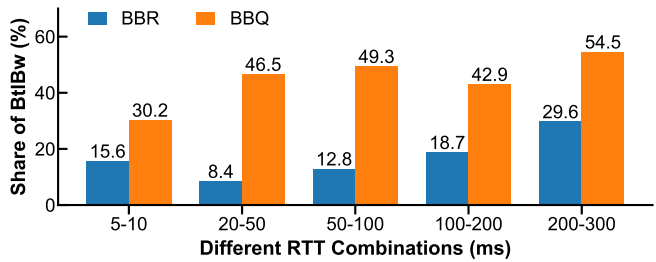}
	\caption{ A small $\alpha$ (3~ms) is a one-size-fits-all choice for typical Internet connections with RTTs ranging from 5~ms to 300~ms. The group is named after the RTTs of the two competing flows. }
	\label{fig:sol-rtts}
\end{figure}
\subsection{A One-Size-Fits-All Cap}
\label{sec:parameter}

We have recommended a small cap $\alpha = 3$~ms for a probing period in
Sec.~\ref{sec:guidelines}. We now show that such a one-size-fits-all
parameter works well for most Internet connections. 
We evaluate the situation where two flows with different RTTs,
ranging from 5~ms to 300~ms, compete on a bottleneck link of 100~Mbps. 
\figref{sol-rtts} compares the bandwidth share of the flow with shorter RTT
using \paper and BBR. In all cases, by setting $\alpha = 3$~ms, \paper consistently
results in a fairer allocation, improving the bandwidth share
of the short flow by at least 84.12\% (200-300~ms) and by up to 4.4$\times$
(20-50~ms) as compared with BBR.


\section{Related Work}
\label{sec:related}

Loss-based congestion control favors short RTT flows \cite{lakshman1997, brown2000}. Flows using the traditional AIMD algorithm \cite{jacobson1988congestion, newreno, dctcp-sigmetrics} increase their window size by one packet per RTT. As a result, short RTT flows grab bandwidth more quickly and settle on a higher \texttt{CWND} in the steady state. By making the window control function independent of RTT, CUBIC \cite{bic, ha2008cubic} achieves linear RTT fairness, meaning, the throughput achieved by a TCP flow is inversely proportional to its RTT. 
Delay-based congestion control \cite{vegas, compoundtcp, timely} uses network latency as a signal of congestion. Those algorithms also favor short RTT flows, because the increase of window size \cite{vegas} or the adjustment of transmission rate \cite{timely} is still inversely proportional to RTT.

BBR \cite{cardwell2017bbr} is a new type of congestion control that proactively models a TCP connection, and paces its rate based on the estimate of the maximum bottleneck bandwidth and minimum RTT. Unlike loss- and delay-based congestion control, BBR presents an opposite bias \emph{against} short RTT flows.
To our knowledge, this paper is the first to identify and quantify such a 
severe RTT fairness problem for BBR.
Our in-depth analysis reveals the root cause of BBR's bias against short RTT flows, based on which we propose our solution, \paper.

The RTT fairness problem can also be addressed by deploying fair queueing
algorithms \cite{drr} in switches, provided that the per-flow queues are 
available \cite{codel}. Unfortunately, switches in the Internet usually lack support for such fine-grained functionalities. Moreover, it is expensive to constantly tweak the fairness rules on all the switches to adapt to dynamic traffic. \paper is a pure end-based solution, and can work with AQM for more diversified performance requirements.
\section{Conclusion}

In this paper, we systematically analyzed the extent and cause of BBR's
RTT fairness problem through extensive measurement studies. We confirmed
that a BBR flow with longer RTT dominates a flow with shorter RTT,
irrespective of network configurations. We showed, through a deep dive
into the flow behaviors, that such a bias is introduced in two phases: (1)
when BBR flows probe for more bandwidth, a persistent queue forms on the
bottleneck due to the excess traffic; (2) the long RTT flow probes for
longer time than the short RTT flow, and hence floods in more excess
traffic, dominating the queue backlog while overwhelming the short RTT
flow. Based on these findings, we designed and implemented BBQ that can
provide significantly better RTT fairness without compromising full
delivery rate or low latency. BBQ constantly detects an excess queue,
and when a queue forms, it enforces a short period of time for probing, so
as to refrain long RTT flows from pouring too much excess traffics into
the pipe. Evaluation shows that BBQ outperforms BBR with respect to RTT
fairness by up to 6.1$\times$, and achieves the same link utilization with even
shorter end-to-end latency.

\bibliographystyle{IEEEtran}
\bibliography{main}
\end{document}